\documentclass[a4paper]{jpconf}
\usepackage{graphicx}

\newcommand\ro{{\hat{\rho}}}

\newcommand\Ho{\hat H}
\newcommand\qo{\hat q}
\newcommand\qoav{\langle\qo\rangle}
\newcommand\half{{\scriptstyle\frac{1}{2}}}
\newcommand\fo{\hat f}
\newcommand\rb{\mathbf{r}}
\newcommand\foavr{\langle\fo^\sigma(\rb)\rangle}
\newcommand\Sc{\mathcal{S}}
\newcommand\Mc{\mathcal{M}}

\begin{document}
\title{Is spontaneous wave function collapse testable at all?}

\author{Lajos Di\'osi}

\address{Wigner Research Centre for Physics, H-1525 Budapest 114, P.O.Box 49, Hungary}

\ead{diosi.lajos@wigner.mta.hu}

\begin{abstract}
Mainstream literature on spontaneous wave function collapse never reflects on or profit from
the formal coincidence and conceptual relationship with standard collapse under time-continuous 
quantum measurement (monitoring). I propose some easy lessons of standard monitoring theory
which would make spontaneous collapse models revise some of their claims. In particular,
the objective detection of spontaneous collapse remains impossible as long as the correct
identification of what corresponds to the signal in standard monitoring is missing from
spontaneous collapse models, the physical detectability of the ``signal'' is not stated explicitly
and, finally, the principles of physical  detection are not revealed.
\end{abstract}

\section{Introduction}
Spontaneous wave function models \cite{GRW86,Bel87,Dio89,GhiPeaRim90,GhiGraBen95}, 
reviewed by \cite{BasGhi03,Basetal13}, dynamically violate the superposition  principle of 
quantum mechanics, assuming tiny spontaneous time-continuous collapse of the 
wave function. For massive degrees of freedom spontaneous collapse gets amplified
and will result in classical behaviour in the objective way. The toolkit of standard quantum 
measurements is no more requested, it is always replaced by spontaneous collapses.
Since collapse and classicality only appear at the level of the mathematical formalism,
additional considerations are used to identify which mathematical object of the given
spontaneous collapse model should represent the emerged classical entities.  

I complain about  mainstream literature on spontaneous collapse for it ignores the lessons 
of standard collapse. Lessons of time-continuous measurement (monitoring) theory \cite{Car93,WisMil10}
are obligatory and instructive for spontaneous collapse, even if one would not implement them all. 

The present analysis  shows that in current spontaneous collapse models 
the proposed collapse is illusory because it is not testable by objective detection.
The sole testable effect is spontaneous decoherence, i.e., the degradation of certain interference terms.
Mathematical apparatus of spontaneous collapse models is redundant: the stochastic 
Schr\"odinger equation  (SSE) is untestable.  The master equation (ME), governing the
density matrix, does encode all possible objectively testable effects.    

\section{Continuous measurement and collapse in standard quantum mechanics}
Considering the continuous measurement (monitoring) of the position $\qo$
of a quantized particle, I published two Ito stochastic equations in 1988 \cite{Dio88a,Dio88b}. 
A plausible expression yields the measurement outcome (also called signal) $q_t$:
 \begin{equation}\label{signal}
q_t dt=\qoav_t dt+\frac{1}{2\sqrt{D}}dW_t ,
\end{equation}
where $\qoav_t=\langle\psi_t\vert\qo\vert\psi_t\rangle$ is the expectation
value of $\qo$ in the current quantum state $\psi_t$ while $W_t$ is the  standard Wiener process (cf.: white-noise).
The power of the noise is inverse proportional to the parameter $D$ which, as we see below,
controls the precision of monitoring. The  evolution of the state vector $\psi_t$ under
coordinate monitoring is governed by the SSE:
\begin{equation}\label{SSE}
d\psi_t=-\frac{i}{\hbar}\Ho\psi_t dt-
\frac{D}{2\hbar^2}(\qo-\qoav_t)^2\psi_t dt
+\frac{\sqrt{D}}{\hbar}(\qo-\qoav_t)\psi_t dW_t .
\end{equation}
It causes dynamical collapse (localization) of the wave function which is
the natural consequence of position measurements.
The theory of q-monitoring contains \emph{two} equations: (\ref{signal}) and (\ref{SSE}). 

If the signal $q_t$ is not accessible or just not recorded, the SSE becomes redundant. 
It implies the following ME for the  ensemble average (density
matrix)  $\ro_t=\langle\psi_t\psi_t^\dagger\rangle_{st}$ of the stochastic $\psi_t$:
\begin{equation}\label{ME}
\frac{d\ro_t}{dt}=-\frac{i}{\hbar}[\Ho,\ro_t]-\frac{D}{2\hbar^2}[\qo,[\qo,\ro_t]] .
\end{equation}
This ME predicts decoherence only; it does not predict collapse.

It turned out soon that the above equations of continuous measurement (monitoring) 
do follow from standard quantum  theory of collapse 
if we interpret monitoring in terms of infinite frequent repeated 
standard measurements. Quantum monitoring has since 1988 become
a precise discipline of standard quantum theory, an unavoidable 
model of modern experiments, see  monographs \cite{Car93,WisMil10}.

\section{Spontaneous collapse in modified quantum mechanics}
When the above dynamical theory of \emph{real} monitoring and \emph{standard} 
collapse emerged, about the same years, various dynamical models of hypothetical \emph{spontaneous}
collapse emerged (see Sec. 1). My  proposal \cite{Dio89} mimics as if hidden devices
would be monitoring the whole Universe.
In the introductory model, hidden monitoring concerns all particle positions. 
More physical is the gravity-related model,  where hidden monitoring concerns 
the (non-relativistic) mass density $\fo(\rb)$ at each point $\rb$ of the Universe.
Full analogy with dynamical theory of real monitoring (Sec. 2) is exploited by \cite{Dio90}
in path integral formalism equivalent to Ito's.

However, the rest of the spontaneous collapse models and the mainstream works \cite{BasGhi03,Basetal13}
never reflect on the mathematical coincidence and conceptual relationship with standard quantum monitoring. 
Hence they do not profit from the \emph{lessons} of standard quantum monitoring. 
Wearing such blindfold against these lessons might be an innocent stance but it is not. 

\section{Continuous Spontaneous Localization}
For concreteness, we consider the Continuous Spontaneous Localization
(CSL) model \cite{GhiPeaRim90,GhiGraBen95} which is similar to the gravity-related
model \cite{Dio89} in that it localizes the mass density $\fo(\rb)$ instead of particle positions.
The definitive equation of CSL coincides with the trivial generalization of the SSE  (\ref{SSE}) for 
$\fo(\rb)$---more precisely: for its coarse graining $\fo^\sigma(\rb)$---in place of $\qo$ \cite{foot1}:
\begin{equation}\label{SSECSL}
d\psi_t=-\frac{i}{\hbar}\Ho\psi_t dt-
\frac{\gamma}{2m_0^2}\int d\rb [\fo^\sigma(\rb)-\foavr_t]^2\psi_t dt
+\frac{\sqrt{\gamma}}{m_0}\int d\rb[\fo^\sigma(\rb)-\foavr_t]\psi_t dW_t(\rb)
\end{equation}
where $\sigma=10^{-5}$cm is the standard width of Gaussian course-graining, 
$\gamma$ is a second CSL-specific parameter and $m_0$ is the atomic mass unit. 
The noise $W_t(\rb)$ is a field of spatially uncorrelated standard Wiener processes. 

The definitive equation of CSL, when used to evolve the density matrix,
yields the following closed linear ME:
\begin{equation}\label{MECSL}
\frac{d\ro_t}{dt}=-\frac{i}{\hbar}[\Ho,\ro_t]-
\frac{\gamma}{2m_0^2}\int d\rb [\fo^\sigma(\rb),[\fo^\sigma(\rb),\ro_t]] .
\end{equation}
  
There is a tendency to attribute physical
status to the noise field $W_t(\rb)$ but the mainstream judgement admits that ``at this stage this is only speculation.
[...] one has [yet] to justify the non-Hermitian coupling and the nonlinear character of the collapse equations'' \cite{Basetal13}.
CSL is a little perplexed by the missing interpretation of the noise. 

Four helpful lessons of standard monitoring, relevant for CSL, follow.

\subsection{Lesson I: Noise is measurement signal noise}
Standard quantum mechanics offers the following unique interpretation for the CSL noise.
CSL corresponds to standard monitoring (\ref{signal}-\ref{ME})---by hidden devices 
this time---of  the mass density $\fo^\sigma(\rb)$ at all locations $\rb$. 
The signal equation (\ref{signal}) would then read
\begin{equation}\label{signalCSL}
\Sc(\rb,t)=\foavr_t +\frac{m_0}{2\sqrt{\gamma}}w_t(\rb) ,
\end{equation}
were the Ito differentials $dW_t(\rb)$ have formally been replaced by $w_t(\rb)dt$ and $w_t(\rb)$ are independent standard white-noises for all $\rb$.
The natural interpretation of CSL noise $W_t(\rb)$
is obvious: it is signal noise where the signal $\Sc(\rb,t)$ is the classical outcome of monitoring $\fo^\sigma(\rb)$.
However, CSL turns blind eyes toward $\Sc(\rb,t)$ and its equation (\ref{signalCSL}).

\subsection{Lesson II: Signal is the only tangible variable}
It is correctly felt in the literature \cite{BasGhi03,Basetal13} that a complete CSL model should specify 
what unique classical  configurations the  basic SSE (\ref{SSECSL})
is to describe. The preferred choice  \cite{BasGhi03} is that it is the quantum average of ($\sigma$-smeared) mass density: 
\begin{equation}\label{avsignalCSL}
\Mc(\rb,t)=\foavr_t =\langle\psi_t\vert\fo^\sigma(\rb)\vert\psi_t\rangle .
\end{equation}
It is thought, in particular, that this mean-field ``is accessible at the macrolevel'' and
``behaves in a classical way'' \cite{BasGhi03}.  Unfortunately, this is not really so.

If we compare $\Mc(\rb,t)$ to the signal (\ref{signalCSL}), we observe that the signal contains a noise term as well:
\begin{equation}
\Sc(\rb,t)=\Mc(\rb,t)+\frac{m_0}{2\sqrt{\gamma}}w_t .
\end{equation}
The role of the noise term is crucial if we desire that the mass distribution in question be common sense
classical field. I proposed the term \emph{tangible} for such variables (fields) because they can be coupled at will
to other fields, including that they can be used at will to control any feed-back on the quantum system itself.
This is why I talked about ``Free Will Test'' (FWT) of tangibility \cite{Dio12}.
Since the signal $\Sc(\rb,t)$ is nothing else just a sequence of standard quantum measurement
outcomes, it passes the FWT. Let us, for instance, modify the Hamiltonian by a simple feed-back term:
\begin{equation}\label{HFBCSLsignal}
\Ho+g\int d\rb\fo^\sigma(\rb) \Sc(\rb,t) .
\end{equation}
Substituting (\ref{signalCSL}) for $\Sc(\rb,t)$ and the above Hamiltonian for $\Ho$ in the SSE (\ref{SSECSL}),
we get a modified SSE such that for $\ro_t$ the ME (\ref{MECSL}) does survive with modified Hamiltonian and modified decoherence coefficient, resp.: 
\begin{equation}\label{MECSLfb}
\Ho\Rightarrow\Ho+\half g\int d\rb [\fo^\sigma(\rb)]^2,~~~~
\frac{\gamma}{m_0^2}\Rightarrow\frac{\gamma}{m_0^2}+\frac{g^2}{4\hbar^2}\frac{m_0^2}{\gamma} .
\end{equation}
(To reconstruct the derivation, works \cite{Dio90},\cite{DioGis95} or \cite{WisMil10} may be studied.)  
If, however, we complete the SSE (\ref{SSECSL}) by a feed-back controlled by $\Mc(\rb,t)$ instead of $\Sc(\rb,t)$:
\begin{equation}\label{HFBCSLmean}
\Ho+g\int d\rb\fo^\sigma(\rb) \Mc(\rb,t) ,
\end{equation}
then the modified SSE will no more allow for any closed linear evolution of $\ro_t$.
Any feed-back control variable, different from the signal  (or functional of it) will 
jeopardize the autonomous linear equation for $\ro_t$ whose loss means loss of consitency \cite{Dio88c,Gis89,Gis90,GisRig95}.
That makes the feed-back (\ref{HFBCSLmean}) illegitimate, hence  mean-field mass density $\Mc(\rb,t)$ is not tangible,
it does not  behave in a ``classical way''.

Of course the mean-field  $\Mc(\rb,t)$ will approximate the signal $\Sc(\rb,t)$ if the noise
on the right-hand-side of (\ref{signalCSL}) is averaged out. Therefore $\Mc(\rb,t)$
might approximate the predicted classical configuration if we suitably impose a time-average on it.   

\subsection{Lesson III: Bell chooses tangible variables}
The GRW jump model of spontaneous collapse \cite{GRW86}, whose diffusive mass-proportional limit is CSL, could also be
identified as position monitoring of constituents by randomly fired standard (though hidden) von Neumann
detectors \cite{Dio00}. When in 1987 John Bell casts the GRW model into its ultimate form,
he also asks for the ``mathematical counterparts in the theory to real events at definite places and times in the real world'' \cite{Bel87}.
His choice is those space-time points, later called flashes (cf., e.g. \cite{Alletal08} and references therein), where GRW jumps are being centred. 
These  flashes correspond exactly what the said (hidden) von Neumann detectors would record as measurement outcomes,
hence flashes are perfect classical (tangible) entities, they pass the FWT. 
Although Bell does not mention the resemblance of GRW to standard monitoring (by hidden detectors) his intuition works perfectly.
Subsequent works on CSL ignore the fact that the signal (\ref{signalCSL}) would be nothing else than the diffusive limit of flashes
of a GRW-like jump model (where random jumps localize on the field $\fo^\sigma(\rb)$ instead of positions). The contrary  is believed: 
mean-field matter density  (\ref{avsignalCSL}) ``must be taken because it [CSL] does not work with flashes''  \cite{FelTum12}.
I think CSL does work with flashes which are just the signal (\ref{signalCSL}).  

\subsection{Lesson IV: SSE is empirically redundant}
 In standard theory of monitoring  and collapse (Sec. 2) it is crucial for the empirical
 testability of the state vector $\psi_t$ under monitoring that the signal $q_t$ be accessible and recorded.
If it is not accessible for some technical reasons, or it is just not recorded, then the behaviour of the state vector
$\psi_t$ is irrelevant because it is not testable/tested empirically. All possible testable predictions are encoded in the
average state, i.e., in the density matrix $\ro_t$. In this case the SSE (\ref{SSE}) becomes redundant,
the ME (\ref{ME}) yields all testable/tested effects which consist of decoherence, as we already said.

The above lesson from standard monitoring, if applied to CSL, tells us the following. 
Since CSL does not interpret the signal at all, it remains empirically unaccessible, it can of course not be recorded
either. Therefore the CSL stochastic wave function $\psi_t$ is empirically irrelevant: all testable predictions of
CSL are encoded in the density matrix $\ro_t$ and its ME (\ref{MECSL}). These testable predictions consist of
decoherence. Spontaneous collapse is not testable at all, the SSE (\ref{SSECSL}) is physically redundant.

\section{Closing remarks}
It is obvious that my criticising the choice of mean density $\Mc(\rb,t)$ for classical predictions is irrelevant
as long as its detection is meant by \emph{subjective perception}. That $\Mc(\rb,t)$ is not tangible (fails FWT) becomes 
relevant only when the goal is {\emph{objective detection}, typically by \emph{coupling} $\Mc(\rb,t)$ to a device or elsewhere.

It is also obvious that spontaneous collapse models differ from standard quantum monitoring, they may well
differ in much more than ``hiddennes'' of the fictitious monitoring devices, especially because we assume a lot of
freedom in how we wish to interpret spontaneous collapse. My work warns that such freedom may
not be that much as believed. 

The statements that  the SSE is redundant, the ME is sufficient, spontaneous decoherence is testable, spontaneous
collapse is not, appeared already in 1989 \cite{Dio89}. These statements hold for all models of spontaneous collapse. 
Detailed arguments were given for CSL, just for concreteness. Independently of the power (even of validity)
of my arguments, let me formulate the central claim in the practical way. Apparently, any \emph{objective} detection 
proposed so far turns out to test decoherence effects fully  calculable via the corresponding master equation.
Therefore, apparently, the promise of spontaneous collapse theories to objectify collapse does not fulfils yet. 

\ack
Support by the Hungarian Scientific Research Fund under Grant No. 103917
and by the EU COST Actions MP1006, MP1209 are acknowledged.

\end{document}